# Enhanced electrocaloric and energy-storage properties of environment-friendly ferroelectric $Ba_{0.9}Sr_{0.1}Ti_{1-x}Sn_xO_3$ ceramics

H. Zaitouni[1,*], L. Hajji[1], D. Mezzane[1,2], E. Choukri[1], A.G. Razumnaya[2,3], Y. Gagou[2], K. Hoummada[4], A. Alimoussa[1], B. Rožič[5], D. Črešnar[5], M. El Marssi[2], Z. Kutnjak[5]

[1]IMED-Lab, Cadi-Ayyad University, Faculty of Sciences and Technology, Department of Applied Physics, Marrakech, Morocco.

[2]Laboratory of Physics of Condensed Matter (LPMC), University of Picardie Jules Verne, Scientific Pole, 33 rue Saint-Leu, 80039 Amiens Cedex 1, France.

[3]Faculty of Physics, Southern Federal University, Rostov-on-Don, 344090, Russia

[4]Laboratoire Matériaux et Microélectronique Nanosciences de Provence, Aix Marseille Université-CNRS, Faculté des Sciences, Marseille, France.

[5]Laboratory for calorimetry and dielectric spectroscopy, Condensed Matter Physics Department, Jozef Stefan Institute, Ljubljana, Slovenia.

## Abstract

The electrocaloric (EC) effect and energy storage properties of eco-friendly ferroelectric $Ba_{0.9}Sr_{0.1}Ti_{1-x}Sn_xO_3$ (BSTS-x) ceramics prepared by the conventional solid-state reaction method were studied. Significant energy efficiency exceeding 80% was found in our samples. In particular, BSTS-5 and BSTS-10 samples exhibit 92% and 88% efficiency, respectively, over a wide temperature range around room temperature (RT). Direct EC measurement and an indirect method based on the thermodynamic approach were used to characterize the EC effect, and both yielded consistent results. The largest electrocaloric responsivity of 0.73 K mm/kV was obtained for BSTS-0 at 368 K with an adiabatic temperature change $\Delta T_{EC}$ of 0.55 K at a low applied electric field of only 7.4 kV/cm. With increasing Sn-content, the EC response of BSTS-x ceramics decreases, while their diffuseness increases over an extensive temperature range, centered at RT. A significant coefficient of performance (COP > 26) was found for our samples. The obtained results demonstrate the possibility of designing eco-friendly materials with higher EC effect and energy efficiency for potential solid-state refrigeration and energy storage devices.

* Corresponding author: hajar.zaitouni@ced.uca.ma ; hajarzaitouni88@gmail.com

## 1. Introduction

Search for high-efficiency, low-cost and environmentally friendly alternatives to conventional cooling systems has generated a growing interest in the electrocaloric (EC) effect [1, 2]. The EC effect is a physical phenomenon in which a polar material exhibits a reversible temperature change ($\Delta T_{EC}$) and entropy ($\Delta S_{EC}$) when an external electric field (E) is applied or removed under adiabatic conditions [3–5]. In addition, the electrocaloric effect is usually evaluated in terms of electrocaloric responsivity ($\Delta T_{EC}/\Delta E$). The EC effect was intensively studied in the 1960s to 1970s but was not exploited commercially because the reported EC effect values were small [6–8]. Since Mischenko *et al.* [9] reported a giant EC effect value ($\Delta T_{EC}$ = 12 K) in the $Pb(Zr_{0.95}Ti_{0.05})O_3$ ferroelectric thin film, the research on the electrocaloric effect of ferroelectric materials has sparked a new wave of interest [4, 10–12]. Abundant data on the EC effect have been reported in various ferroelectric thin and thick films, polymers, single crystals and bulk ceramics [4, 5, 9, 13]. For ferroelectrics, a large entropy change associated with a polarization change is critical to have a large EC effect. Conventional lead-based ferroelectric materials are used in this direction [14, 15]. Nevertheless, these lead-based electrocaloric materials suffer from environmental concerns. In this context, various lead-free refrigeration materials have been investigated over the years, e.g., $BaTiO_3$ [4, 16], $SrTiO_3$ [17], $Na_{0.5}Bi_{0.5}TiO_3$ [18].

Among all, lead-free $BaTiO_3$ (BT) based compounds have been extensively studied as an environmentally friendly material that exhibits excellent dielectric, ferroelectric and piezoelectric properties [19–21]. Recently, BT has been regarded as one of the potentially promising lead-free ceramics for future solid-state refrigeration applications [4]. As well known, a large EC effect usually occurs in the vicinity of the ferroelectric–paraelectric (FE – PE) phase transition. Thus, to use BT as an EC material in refrigeration technologies, the Curie temperature ($T_C$) must be shifted much lower than its value (~120 °C) in pure BT ceramic. In this direction, doping at A and/or B-sites of the $BaTiO_3$ matrix can efficiently raise its $T_C$ to room temperature, which may simultaneously enhance its EC effect [22]. For example, a significant EC effect ($\Delta T_{EC}$ = 4.5 K, $\Delta T_{EC}/\Delta E$ = 0.31 K mm/kV under 145 kV/cm,) at 312 K was given by Qian *et al.* [23] in $Ba(Zr_{0.2}Ti_{0.8})O_3$ ceramics. Sanlialp *et al.* [24] also found a high EC temperature change of 0.63 K at 317 K under a low electric field of 20 kV/cm in $BaTi_{0.89}Sn_{0.11}O_3$ ceramics. Lately, Lu et al. [25] reported a large EC effect strength of $\Delta T_{EC}/\Delta E$ = 0.52 K mm/kV at 309 K in $Ba_{1-x}Sr_xSn_{0.05}Ti_{0.95}O_3$ (x = 0.10) ceramics. In addition, our previous work [26] has demonstrated that the partial substitution of $Sn^{4+}$ in $Ba_{0.9}Sr_{0.1}TiO_3$ (BST) lattice ($Ba_{0.9}Sr_{0.1}Ti_{1-x}Sn_xO_3$, x = 0.00 – 0.15) shifted the phase transition to room temperature (RT) and also improved its dielectric and ferroelectric properties. Thus, a part of the present paper focuses on the lead-free ferroelectric $Ba_{0.9}Sr_{0.1}Ti_{1-x}Sn_xO_3$ (denoted BSTS-x) ceramics as potential candidates for solid-state refrigeration applications. It is worth noting that a detailed description of the crystal structure, the microstructure and tuning properties of the BSTS-x samples have been reported elsewhere [26].

Besides solid-state cooling applications, BT-based materials have also demonstrated potential applications for energy storage capacitors [27–31]. Indeed, when compared to batteries and electromechanical capacitors, it has been shown that these systems can provide high energy density with an ultra-fast charge/discharge process and a long lifetime. As a result, they are very appealing for advanced electronic applications such as electric power systems, high-frequency inverters and power grids [32, 33].

According to several studies, dielectric materials with large saturation polarization ($P_{max}$), low remnant polarization ($P_r$), and low dielectric loss, particularly antiferroelectric and relaxor ferroelectric materials, are thought to have excellent energy storage performances [34–36]. In the case of $Ba_{0.9}Sr_{0.1}Ti_{1-x}Sn_xO_3$ solid-solutions, varying the Sn-content in the B-site of the perovskite BST promotes the diffuse phase transition, and thus the relaxor behavior can provide higher energy storage properties; i.e., the energy storage density ($W_{rec}$) and the energy storage efficiency ($\eta$).

In this study, the EC temperature change of BSTS-x ceramics was calculated using Maxwell's relation and the P – E hysteresis curves obtained at different temperatures (indirect method) and directly measured using a high-resolution calorimeter. The results procured by both approaches will be discussed and compared. Furthermore, the EC responsivity, the coefficient of performance (COP) and electrical storage energy performances have been determined and discussed to understand better the versatile properties of the materials towards electronic systems for solid-state refrigeration and energy storage applications.

## 2. Experimental procedure

Polycrystalline BSTS-x ceramics with nominal compositions x = 0.00, 0.05 and 0.10 were fabricated via the conventional solid-state reaction method using barium carbonate ($BaCO_3$, 99.8%), strontium carbonate ($SrCO_3$, 99.8%), titanium dioxide ($TiO_2$, 99.8%) and stannic dioxide ($SnO_2$, 99.9%) as starting raw materials from Alfa Aesar company. A detailed description of sample preparation and phase diagram can be found elsewhere [26]. Before performing the measurements, silver electrodes were coated on both sides of the ceramic samples. Temperature-dependent P–E hysteresis loops were recorded at 5 Hz using an aixACCT, TF Analyser 3000 ferroelectric test system.

The direct electrocaloric measurements were carried out using a high-resolution calorimeter allowing precise temperature stabilization of the bath (within 0.1 mK) and high-resolution measurements of the sample temperature variation. The temperature was measured via a small bead thermistor. More details on direct EC measurement setup are reported elsewhere [37, 38].

## 3. Results and discussion

### 3.1. Dielectric analysis

In our previous paper [26], the results of dielectric measurements showed that with increasing Sn-content in BST lattice, the three transitions merge into one broad peak near RT at x = 0.10, i.e., at the quasi-quadruple point in the phase diagram (inset of Fig. 1). As a consequence, the dielectric constant is highly improved. This enhancement was explained by Shi *et al.* for Sn doped BT [39], as discussed in our earlier work [26]. They have noticed that the formed Polar Nano-Regions (PNRs) in the matrix BT under matrix stress, which should have smaller lattice constants due to their lower Sn concentration, expand. As a result, the dielectric constant values are improved significantly [39].

In order to precisely determine the diffuseness dielectric curves degree, we used the equation proposed by Santos–Eiras [40] given as follow:

$$\varepsilon' = \frac{\varepsilon'_m}{1+\left(\frac{T-T_m}{\Delta}\right)^{\gamma}}, \qquad (1)$$

where $\varepsilon'_m$ is the maximum dielectric constant at the transition temperature $T_m$, the exponent γ reflects the character of the ferroelectric–paraelectric (FE–PE) phase transition and Δ is the peak broadening which defines the diffuseness degree of the FE–PE phase transition. Temperature dependence dielectric constant plots were fitted with Eq. (1) at 1 kHz for the studied BSTS-x samples (Fig. 1). The coincidence between experimental data and the fit validates the applied model. The fitted parameters are given in Table 1.

The obtained value of $\gamma$ close to 1 reveals the character of BSTS-0 as a normal ferroelectric. In contrast, $\gamma$ = 1.36 and 1.61 were found for BSTS-5 and BSTS-10, respectively, revealing an incomplete ferroelectric phase transition. Furthermore, it is found that the peak broadening values $\Delta$ lie in the range 12.41 – 15.70 and increase with increasing Sn-content. These findings indicate that the phase transitions of BSTS-x solid solution are progressively diffused, and EC effect peaks may broaden in varying degrees. Finally, it should be noted that no pronounced relaxor behavior is detected for our compositions. Indeed, significant B site doping strengthens the relaxor behavior [41]. However, in our case, the Sn amount added into the BST lattice is insufficient to induce relaxor behavior, which is consistent with some studies in the literature [42].

**(Insert Fig. 1, here)**

**(Insert Table 1, here)**

### 3.2. Energy storage properties

As a well-known fact, ferroelectric materials possess the electric energy storage capacity and can be obtained from P–E hysteresis loops. According to the definition of the energy storage density, the recoverable energy density ($W_{rec}$) and the total energy density ($W_{tot}$) are defined as:

$$W_{rec} = \int_{P_r}^{P_{max}} E dP, \quad (2)$$

$$W_{tot} = \int_{0}^{P_{max}} E dP, \quad (3)$$

Based on the above equations (2) and (3), $W_{rec}$ can be estimated by numerical integration of the area between P–E loops discharge curve and polarization axis as schematically depicted in Fig. 2. While $W_{tot}$ can be determined by adding the area gathering both $W_{rec}$ (the green area in Fig. 2) and energy loss density $W_{loss}$ (the red area in Fig. 2).

**(Insert Fig. 2, here)**

To evaluate the energy density stored in our BSTS-x system, we used P–E hysteresis loops measurements shown exemplarily in Fig. 3(a-c). More details on the ferroelectric properties of the studied system are reported elsewhere [26]. The recovered energy and the total energy have been calculated from P–E loops as a function of temperature, and the results are shown in Fig. 4(a) and (b). It can be seen that BSTS-x samples show a maximum value of $W_{rec}$ at the Curie temperature, which is consistent with the dielectric measurements. Indeed, our compositions display modest values of $W_{rec}$ as compared to literature due to the very low applied electric field (< 10 kV/cm). For instance, the BSTS-0 ceramic shows a value of $W_{rec}$ and $W_{tot}$ of 24 mJ/cm$^3$ and 50 mJ/cm$^3$, respectively, at a low applied electric field of 7.4 kV/cm. However, it was demonstrated in previous works on energy storage capacities that a large electric field can induce higher polarization. Consequently, the energy storage properties increase inevitably based on the definition of energy-storage density. Unfortunately, such a large electric field is not required for energy-storage devices [43].

**(Insert Fig. 3, here)**

From practical applications, the energy storage efficiency ($\eta$) is an important parameter for evaluating energy storage performances with no electric field limitation. An expression of $\eta$ is given as follow:

$$\eta = \frac{W_{rec}}{W_{rec}+W_{loss}}, \tag{4}$$

where $W_{rec}$ and $W_{loss}$ are the recoverable energy density and energy loss density, respectively. A small value of $\eta$ indicates significant losses in the form of the hysteresis (red area in Fig. 2). By reducing the energy loss density, high energy storage efficiency can be achieved, as evidently shown in Eq. (4). The energy storage efficiency of BSTS-x samples has been calculated and shown as a function of temperature and Sn-content in Fig. 4 (c) and (d), respectively. It is evident from the bar chart (Fig. 4d) that all the BSTS-x ceramics perform high storage efficiency values exceeding 84%. This is because the weak coercive field and small remanent polarization lead to a very slim hysteresis loop, resulting in lower energy loss density and high storage efficiency. Therefore, BSTS-5 and BSTS-10 show higher energy storage efficiency than the pure BSTS-0, i.e., 92% and 88% near the RT, respectively.

On the other hand, the temperature dependence of energy storage efficiency of BSTS-x samples (Fig. 4c) shows that BSTS-5 and BSTS-10 ceramics exhibit a high storage efficiency above 80% over a wide temperature range of 60 K and 80 K, respectively. These results are consistent with the broad phase transition of BSTS-x samples.

**(Insert Fig. 4, here)**

Table 2 compares the energy storage properties of our BSTS-x samples with other Pb-free ferroelectrics found in the literature. Our results show modest recoverable energy density values than other ferroelectrics due to the low applied electric field (< 8 kV/cm). However, under the same conditions of the electric field, our results display comparable or even stunning values of $W_{rec}$, e.g., $Ba_{0.85}Ca_{0.15}Zr_{0.10}Ti_{0.90}O_3$ ceramics (14 mJ/cm$^3$ at 6.5 kV/cm) [44] or $BaTi_{0.895}Sn_{0.105}O_3$ ceramics (31 mJ/cm$^3$ at 10 kV/cm) [45]. Moreover, regardless of the applied electric field, the BSTS-x ceramics present the highest values of energy storage efficiency compared to the literature data presented in Table 2. Therefore, all these results may provide a promising way to design high energy storage devices with high storage efficiency by controlling the doping concentration in the BSTS-x system.

**(Insert Table 2, here)**

### 3.3. Electrocaloric studies

The indirect Maxwell approach was used to compute the electrocaloric effect. Here, the EC effect is calculated from the measured ferroelectric order parameter P(E,T), extracted from the upper branches of the corresponding P–E hysteresis loops. Fig. 3(d-f) displays the thermal evolution of the polarization (P) at different applied electric fields, which continuously decreases with increasing temperature. By approaching the Curie temperature $T_C$, the P(T) curves exhibit an apparent anomaly indicating that a significant enhancement of the electrocaloric effect would occur around $T_C$. For instance, an anomaly between 360 K and 380 K was noted in the P(T) curve of BSTS-0 ceramic, close to its Curie temperature $T_C = 368$ K.

Based on the well-known Maxwell's relation $(\partial P/\partial T)_E = (\partial S/\partial E)_T$, the electrocaloric temperature change ($\Delta T_{EC}$) and the isothermal entropy change ($\Delta S$) as the applied electric field changes from ($E_1 = 0$) to $E_2$ can be expressed as follows [9, 46].

$$\Delta T_{EC} = -\frac{1}{\rho}\int_{E_1}^{E_2} \frac{T}{C_p}\left(\frac{\partial P}{\partial T}\right)_E dE, \tag{5}$$

$$\Delta S = -\frac{1}{\rho}\int_{E_1}^{E_2} \left(\frac{\partial P}{\partial T}\right)_E dE, \tag{6}$$

where $\rho$ is the bulk density of ceramics which was found to be 5.94 gcm$^{-3}$, 6.00 gcm$^{-3}$ and 6.08 gcm$^{-3}$ for BSTS-0, BSTS-5 and BSTS-10 samples, respectively, as reported in our earlier work [26]. Note that the obtained values were measured based on Archimedes' principle. The specific heat capacity $C_p$ value was taken as 400 JKg$^{-1}$K$^{-1}$ for all BSTS-x ceramics in the studied range temperature as commonly used for Sn doped BT (BTS) in the literature [47]. From Eq. (5), the critical factor in calculating the EC effect is the pyroelectric coefficient $\left(\frac{\partial P}{\partial T}\right)_E$ which can be obtained from seventh-order polynomial fits of P(T) data shown in Fig. 3(d-f).

The thermal evolution of the electrocaloric temperature change ($\Delta T_{EC}$) and the predicted entropy ($\Delta S$) under selected applied electric fields are shown in Fig. 5 and Fig. 6, respectively. One can see that both curves $\Delta T_{EC}$ and $\Delta S$ reach their maximum values near the Curie temperature for each BSTS-x sample. This result revealed that both $\Delta T_{EC}$ and $\Delta S$ maximum value is assigned to the great change in polarization with the temperature near the Curie temperature. As expected, a significant enhancement of EC response with increasing electric field is observed for each composition, coherent with the variation of P–T curves with the electric field. At the maximum value of electric field, the $\Delta T_{EC}$ at the FE–PE transition reaches a maximum value of 0.55 K (at 368 K and 7.4 kV/cm), 0.21 K (at 341 K and 6.8 kV/cm) and 0.10 K (at RT and 4.4 kV/cm) for BSTS-x ceramics with x = 0.00, 0.05 and 0.10, respectively. The BSTS-x solid solutions ceramics with high Sn-content display smaller $\Delta T_{EC}$ than the undoped (BSTS-0) ceramic but show a broader peak in the vicinity of the RT, consistent with the diffuse phase transition behavior of BSTS-x ceramics.

**(Insert Fig. 5, here)**

**(Insert Fig. 6, here)**

The electrocaloric responsivity ($\xi = \Delta T_{EC}/\Delta E$) has been chosen as a suitable parameter to compare the EC effect in different materials since it is independent of the geometrical form and the dimensions of the materials. Therefore, we have gathered some data about the EC effect in typical lead-free and lead-based ferroelectric materials reported by other authors with the obtained data in this work. The comparative results are displayed in Table 3. In this work, we have found a high electrocaloric responsivity value of $\xi$ = 0.73 K mm/kV at 368 K for BSTS-0 ceramic, which is one of the giant EC responsivity values reported for BT-based materials [12, 25, 48]. Notably, the EC responsivity decreases gradually with increasing Sn-content and show a value of 0.30 K mm/kV and 0.23 K mm/kV for BSTS-5 and BSTS-10, respectively, with a broadened peak near RT, which is desired for practical refrigeration applications.

To assess the reliability of the previous indirect EC results, we have performed the direct EC measurements using a high-resolution calorimeter. The results of the direct EC measurements ($\Delta T_{EC}$ vs. T) of the studied samples at various electric fields are plotted in Fig. 7(a-c). Fig. 7(a-c) insets display the comparison between direct and indirect measurements obtained at the same electric field. The evolution of $\Delta T_{EC}$ versus temperature in the direct method showed the same trend as in the indirect approach, i.e., the peak reaches its maximum just above the FE–PE phase transition. As expected, the results are also stunning due to the high electric field. High electrocaloric response of $\Delta T_{EC}$ = 0.52 K, which corresponds to an EC responsivity of $\xi$ = 0.7 K mm/kV at a low applied electric field of 7.4 kV/cm was found in BSTS-0 sample. As the concentration of Sn increases, the electrocaloric temperature change is reduced and show a broadened peak with a maximum value of 0.47 K at an applied electric field of 16 kV/cm ($\xi$ = 0.3

K mm/kV) and 0.20 K at an applied electric field of 8.5 kV/cm ($\xi$ = 0.22 K mm/kV) for BSTS-5 and BSTS-10 samples, respectively.

**(Insert Fig. 7, here)**

By comparing the electrocaloric responses of BSTS-x samples, we noted that BSTS-0 ceramic displays the highest EC response, more than 50% greater than the peak value of BSTS-10 ceramic. Nevertheless, the dielectric constant ($\boldsymbol{\varepsilon'_m}$) measured under zero DC-field (shown in Fig. 1), for BSTS-10, was nearly two times greater than for BSTS-0. The current electrocaloric study demonstrates that the magnitude of the dielectric constant is not directly linked to the enhancement of the EC response in the BSTS-x system. Instead, one must consider the dielectric permittivity's field dependence, which drops significantly as the electric field increases [49], clarifying the dielectric permittivity's non-direct relationship to the EC response value. In particular, it was demonstrated in our previous work [26] that only 1.4 kV/cm of DC-electric field is required to suppress more than 60% of the original value of the dielectric permittivity of BSTS-10 ceramic. One of the fundamental factors determining a stronger EC response is the latent heat (strongly correlated to the discontinuous change of the polarization at the phase transition), which is considerably higher in materials with first-order phase transitions [50, 51]. Therefore, BSTS-5 and BSTS-10 samples with diffuse phase transition show smaller EC values with broader peaks.

In terms of EC temperature change ($\Delta T_{EC}$) and EC responsivity ($\xi$), it is found that both measurements show similar estimations of EC responses (insets of Fig. 7). The good agreement between both methods demonstrates their validity. The obtained EC values using direct measurements are reported in Table 3 and compared to the available literature. Clearly our samples display the highest electrocaloric responsivity values reported in literature compared to other Pb-free ferroelectrics as well as Pb-based ferroelectric, e.g., the $Ba_{0.65}Sr_{0.35}TiO_3$ ($\Delta T_{EC}$ = 0.42 K and $\xi$ = 0.21 K mm/kV under 20 kV/cm) [52], $Ba_{0.94}Ca_{0.06}Ti_{1-x}Sn_xO_3$ ($\Delta T_{EC}$ = 0.59 K and $\xi$ = 0.30 K mm/kV under 20 kV/cm) [48], or also the BT single crystal reported in Moya *et al* [4], which has similar EC strength value as BSTS-0 ceramic for slightly higher temperature ($\xi$ = 0.75 K mm/kV under 12 kV/cm at 402 K).

For practical applications, the material coefficient of performance (COP) is also an essential criterion in evaluating the efficiency of an electrocaloric material. Here, COP is given by the following equation [53]:

$$COP = \left|\frac{Q}{W}\right| = \frac{|\Delta S \times T|}{W_{rec}}, \quad (7)$$

where Q and $W_{rec}$ are the isothermal heat and corresponding recovered density. The COP values obtained of the different samples are illustrated in Fig. 8 as a function of the composition x. We notice that the maximum COP value is obtained near the FE – PE phase transition, where the EC temperature change ($\Delta T_{EC}$) reaches its maximum. A higher value of ~60 is observed for the composition BSTS-0, which decreases to 31 and 26 for BSTS-5 and BSTS-10, respectively. The obtained values are much higher than those reported in several previous works. For instance, Zhang *et al*. [54] found a maximum COP value of 27.4 in lead titanate $(Pb_{0.8}La_{0.2})Ti_{0.2}O_3$ ceramics, and Zhao *et al*. [55] reported a value of 18 in $Pb_{0.97}La_{0.02}(Zr_{0.75}Sn_{0.18}Ti_{0.07})O_3$ antiferroelectric thick films. Moreover, Srikanth *et al*. [13] found a value of 17 in $Ba_{0.7}S_{r0.3}TiO_3$ bulk ceramics at 303 K under an applied electric field of 33 kV/cm. The synthesized BSTS-x ceramics exhibit excellent EC effect properties, i.e., sizeable $\Delta T_{EC}$, $\Delta S$, $\Delta T_{EC}/\Delta E$ and COP values obtained at the low applied electric field. These properties indicate the BSTS-x eco-friendly ceramics as a potential cooling material for novel solid-state refrigeration devices with high cooling efficiency.

**(Insert Fig. 8, here)**

**(Insert Table 3, here)**

## 4. Conclusion

Lead-free BSTS-x ceramics were prepared via conventional solid-state reaction method, and the corresponding electrocaloric effect and electrical energy storage properties were investigated. The results show that BSTS-x ceramics exhibit modest values of recovered energy density ($W_{rec}$) compared to other ferroelectrics due to the low applied electric field (< 8 kV/cm). Nevertheless, under the same conditions of the electric field, our results show comparable or even stunning values. Indeed, all compounds have high energy efficiency ($\eta$ > 80%). In particular, the compositions BSTS-5 and BSTS-10 present efficiency of 92 % and 88%, respectively, over a wide temperature range, centered at RT. Significant electrocaloric response and high refrigeration efficiency (COP) are achieved. Using Maxwell's thermodynamic approach, a maximum value of EC responsivity $\xi = \Delta T_{EC}/\Delta E = 0.73$ K mm/kV at 368 under a low electric field of only 7.4 kV/cm is obtained in BSTS-0. The EC effect maximum decreases gradually by increasing Sn-content, and the corresponding peak broadens over a wide temperature range near the RT. This result is in accordance with the diffuse phase transition nature of BSTS-x ceramics. Moreover, the direct EC measurements results are consistent with indirect measurements. The combination of these results demonstrates that BSTS-x ceramics are highly promising candidates for eco-friendly refrigeration applications near RT and an energy storage material for electronic and prototype apparatus over a wide temperature range.

**Acknowledgements**

The authors are grateful to the European Community for the generous financial support provided through the H2020MSCA-RISE-ENGIMA-778072 project. In addition, the ARRS project J1-9147, program P1-0125, the CNRST Priority Program (PPR15/2015) and the Ministry of Science and Higher Education of the Russian Federation, grant agreement № 075-15-2021-953 are all gratefully acknowledged.

## References

[1] J. F. Scott, "Electrocaloric materials," *Annu. Rev. Mater. Res.*, vol. 41, pp. 229–240, 2011, doi: 10.1146/annurev-matsci-062910-100341.

[2] T. Correia and Q. Zhang, *Electrocaloric Materials New generation of coolers*. 2014.

[3] M. Valant, "Electrocaloric materials for future solid-state refrigeration technologies," *Prog. Mater. Sci.*, vol. 57, no. 6, pp. 980–1009, 2012, doi: 10.1016/j.pmatsci.2012.02.001.

[4] X. Moya *et al.*, "Giant electrocaloric strength in single-crystal $BaTiO_3$," *Adv. Mater.*, vol. 25, no. 9, pp. 1360–1365, 2013, doi: 10.1002/adma.201203823.

[5] B. Rožič, Z. Kutnjak, B. Neese, S.-G. Lu, and Q. M. Zhang, "Electrocaloric effect in the relaxor ferroelectric polymer composition P(VDF–TrFE–CFE)0.90–P(VDF–CTFE)0.10," *Phase Transitions*, vol. 83, pp. 819–823, 2010, doi: 10.1080/01411594.2010.509145.

[6] P. D. Thacher, "Electrocaloric effects in some ferroelectric and antiferroelectric Pb(Zr, Ti)$O_3$ compounds," *J. Appl. Phys.*, vol. 39, pp. 1996–2002, 1968, doi: 10.1063/1.1656478.

[7] W. N. Lawless, "Specific heat and electrocaloric properties of $KTaO_3$ at low temperatures," *Phys. Rev. B*, vol. 16, no. 1, pp. 433–439, 1977, [Online]. Available: https://link.aps.org/doi/10.1103/PhysRevB.16.433.

[8] B. A. Tuttle and D. A. Payne, "The Effects Of Microstructure On The Electrocaloric Properties Of Pb(Zr,Sn,Ti)$0_3$ Ceramics," *Ferroelectrics*, vol. 37, no. 1, pp. 603–606, 1981, doi: 10.1080/00150198108223496.

[9] A. S. Mischenko, Q. Zhang, J. F. Scott, R. W. Whatmore, and N. D. Mathur, "Giant electrocaloric effect in thin-film $PbZr_{0.95}Ti_{0.05}O_3$," *Science (80-. ).*, vol. 311, pp. 1270–1271, 2006, doi: 10.1126/science.1123811.

[10] G. Dai *et al.*, "Direct and indirect measurement of large electrocaloric effect in barium strontium titanate ceramics," *Int. J. Appl. Ceram. Technol.*, vol. 00, pp. 1–8, 2019, doi: 10.1111/ijac.13384.

[11] S. Qi, G. Zhang, L. Duan, T. Zeng, and J. Cao, "Electrocaloric effect in Pb-free Sr-doped $BaTi_{0.9}Sn_{0.1}O_3$ ceramics," *Mater. Res. Bull.*, vol. 91, pp. 31–35, 2017, doi:

10.1016/j.materresbull.2017.03.026.

[12] Z. Lv, J. Wei, T. Yang, Z. Sun, and Z. Xu, "Manipulation of Curie temperature and ferroelectric polarization for large electrocaloric strength in $BaTiO_3$-based ceramics," *Ceram. Int.*, vol. 46, no. 10, pp. 14978–14984, 2020, doi: 10.1016/j.ceramint.2020.03.027.

[13] K. S. Srikanth, S. Patel, and R. Vaish, "Electrocaloric behavior and temperature dependent scaling of dynamic hysteresis of $Ba_xSr_{1-x}TiO_3$ (x= 0.7, 0.8 and 0.9) bulk ceramics," *J. Aust. Ceram. Soc.*, vol. 3, 2017.

[14] D. Saranya, A. R. Chaudhuri, J. Parui, and S. B. Krupanidhi, "Electrocaloric effect of PMN-PT thin films near morphotropic phase boundary," *Bull. Mater. Sci.*, vol. 32, no. 3, pp. 259–262, 2009, doi: 10.1007/s12034-009-0039-3.

[15] B. Rožič, B. Malič, H. Uršič, J. Holc, M. Kosec, and Z. Kutnjak, "Direct Measurements of the Electrocaloric Effect in Bulk $PbMg_{1/3}Nb_{2/3}O_3$ (PMN) Ceramics," *Ferroelectrics*, vol. 421, no. 1, pp. 103–107, 2011, doi: 10.1080/00150193.2011.594742.

[16] K. S. Srikanth and R. Vaish, "Enhanced electrocaloric, pyroelectric and energy storage performance of $BaCe_xTi_{1-x}O_3$ ceramics," *J. Eur. Ceram. Soc.*, vol. 37, no. 13, pp. 3927–3933, 2017, doi: 10.1016/j.jeurceramsoc.2017.04.058.

[17] W. N. Lawless and A. J. Morrow, "Specific heat and electrocaloric properties of a $SrTiO_3$ ceramic at low temperatures," *Ferroelectrics*, vol. 15, no. 1, pp. 159–165, 1977, doi: 10.1080/00150197708237810.

[18] M. E. Rogers, C. M. Fancher, and J. E. Blendell, "Domain evolution in lead-free thin film piezoelectric ceramics," *J. Appl. Phys.*, vol. 112, no. 052014, pp. 1–5, 2012, doi: 10.1063/1.4746088.

[19] S. Chihaoui, L. Seveyrat, V. Perrin, I. Kallel, L. Lebrun, and H. Khemakhem, "Structural evolution and electrical characteristics of Sn-doped $Ba_{0.8}Sr_{0.2}TiO_3$ ceramics," *Ceram. Int.*, vol. 43, pp. 427–432, 2017, doi: 10.1016/j.ceramint.2016.09.176.

[20] M. Chen *et al.*, "Polymorphic phase transition and enhanced piezoelectric properties in $(Ba_{0.9}Ca_{0.1})(Ti_{1-x}Sn_x)O_3$ lead-free ceramics," *Mater. Lett.*, vol. 97, pp. 86–89, 2013, doi: 10.1016/j.matlet.2012.12.067.

[21] Y. Yang, Y. Zhou, J. Ren, Q. Zheng, K. H. Lam, and D. Lin, "Coexistence of three ferroelectric phases and enhanced piezoelectric properties in BaTiO3–CaHfO3 lead-free ceramics," *J. Eur. Ceram. Soc.*, vol. 38, no. 2, pp. 557–566, 2018, doi: 10.1016/j.jeurceramsoc.2017.09.023.

[22] Z. Luo *et al.*, "Enhanced electrocaloric effect in lead-free $BaTi_{1-x}Sn_xO_3$ ceramics near room temperature," *Appl. Phys. Lett.*, vol. 105, no. 102904, pp. 1–5, 2014, doi: 10.1063/1.4895615.

[23] X. S. Qian *et al.*, "Giant electrocaloric response over a broad temperature range in modified $BaTiO_3$ Ceramics," *Adv. Funct. Mater.*, vol. 24, no. 9, pp. 1300–1305, 2014, doi: 10.1002/adfm.201302386.

[24] M. Sanlialp *et al.*, "Direct measurement of electrocaloric effect in lead-free $Ba(Sn_xTi_{1-x})O_3$ ceramics," *Appl. Phys. Lett.*, vol. 111, no. 173903, pp. 1–6, 2017, doi: 10.1063/1.5001196.

[25] S. G. Lu *et al.*, "Enhanced electrocaloric strengths at room temperature in $(Sr_xBa_{1-x})(Sn_{0.05}Ti_{0.95})O_3$ lead-free ceramics," *J. Alloys Compd.*, vol. 871, no. 159519, pp. 1–9, 2021, doi: 10.1016/j.jallcom.2021.159519.

[26] H. Zaitouni *et al.*, "Structural, dielectric, ferroelectric and tuning properties of Pb-free ferroelectric $Ba_{0.9}Sr_{0.1}Ti_{1-x}Sn_xO_3$," *Ceram. Int.*, vol. 46, no. 17, pp. 27275–27282, 2020, doi: 10.1016/j.ceramint.2020.07.212.

[27] Y. H. Huang, B. Liu, J. Li, and Y. J. Wu, "Improved energy storage performance of $Ba_{0.4}Sr_{0.6}TiO_3$ nanocrystalline ceramics prepared by using oxalate co-precipitation and spark plasma sintering," *Mater. Res. Bull.*, vol. 113, no. January, pp. 141–145, 2019, doi: 10.1016/j.materresbull.2019.01.029.

[28] Z. Xu, H. Qiang, and Y. Chen, "Improved energy storage properties of Mn and Y co-doped BST films," *Mater. Lett.*, vol. 259, no. 126894, pp. 1–3, 2020, doi: 10.1016/j.matlet.2019.126894.

[29] W. B. Li, D. Zhou, and L. X. Pang, "Enhanced energy storage density by inducing defect dipoles in lead free relaxor ferroelectric $BaTiO_3$-based ceramics," *Appl. Phys. Lett.*, vol. 110, no. 132902, pp. 1–5, 2017, doi: 10.1063/1.4979467.

[30] Y. Li *et al.*, "Energy storage performance of $BaTiO_3$-based relaxor ferroelectric ceramics prepared through a two-step process," *Chem. Eng. J.*, vol. 419, no. 129673, pp. 1–12, 2021, doi: 10.1016/j.cej.2021.129673.

[31] J. Lv *et al.*, "Enhancing the dielectric and energy storage properties of lead-free $Na_{0.5}Bi_{0.5}TiO_3$–$BaTiO_3$ ceramics by adding $K_{0.5}Na_{0.5}NbO_3$ ferroelectric," *Ceram. Int.*, vol. 48, no. 1, pp. 22–31, 2022, doi: 10.1016/j.ceramint.2021.08.049.

[32] C. Liu, F. Li, M. Lai-Peng, and H. M. Cheng, "Advanced materials for energy storage," *Adv. Mater.*, vol. 22, no. 8, pp. 28–62, 2010, doi: 10.1002/adma.200903328.

[33] X. Zhang *et al.*, "Giant Energy Density and Improved Discharge Efficiency of Solution-Processed Polymer Nanocomposites for Dielectric Energy Storage," *Adv. Mater.*, vol. 28, no. 10, pp. 2055–2061, 2016, doi: 10.1002/adma.201503881.

[34] Y. Lin *et al.*, "Excellent Energy-Storage Properties Achieved in $BaTiO_3$-Based Lead-Free Relaxor Ferroelectric Ceramics via Domain Engineering on the Nanoscale," *ACS Appl. Mater. Interfaces*, vol. 11, pp. 36824–36830, 2019, doi: 10.1021/acsami.9b10819.

[35] W. Cao *et al.*, "Defect dipole induced large recoverable strain and high energy-storage density in lead-free $Na_{0.5}Bi_{0.5}TiO_3$-based systems," *Appl. Phys. Lett.*, vol. 108, no. 202902, pp. 1–5, 2016, doi: 10.1063/1.4950974.

[36] Q. Yuan, F. Yao, Y. Wang, R. Ma, and H. Wang, "Relaxor ferroelectric $0.9BaTiO_3$-$0.1Bi(Zn_{0.5}Zr_{0.5})O_3$ ceramic capacitors with high energy density and temperature stable energy storage properties," *J. Mater. Chem. C*, vol. 5, no. 37, pp. 9552–9558, 2017, doi: 10.1039/c7tc02478a.

[37] B. Rožič *et al.*, "Direct Measurements of the Giant Electrocaloric Effect in Soft and Solid Ferroelectric Materials," *Ferroelectrics*, vol. 405, no. 1, pp. 26–31, 2010, doi: 10.1080/00150193.2010.482884.

[38] Z. Kutnjak, J. Petzelt, and R. Blinc, "The giant electromechanical response in ferroelectric relaxors as a critical phenomenon," *Nat. Lett.*, vol. 441, no. June, pp. 2–5, 2006, doi: 10.1038/nature04854.

[39] T. Shi, L. Xie, L. Gu, and J. Zhu, "Why Sn doping significantly enhances the dielectric

properties of $Ba(Ti_{1-x}Sn_x)O_3$," *Sci. Rep.*, vol. 5, no. 8606, pp. 1–4, 2015, doi: 10.1038/srep08606.

[40] I. A. Santos and J. A. Eiras, "Phenomenological description of the diffuse phase transition in ferroelectrics," *J. Phys. Condens. MATTER*, vol. 13, pp. 11733–11740, 2001.

[41] M. J. Ansaree, U. Kumar, and S. Upadhyay, "Solid-state synthesis of nano-sized $Ba(Ti_{1-x}Sn_x)O_3$ powders and dielectric properties of corresponding ceramics," *Appl. Phys. A Mater. Sci. Process.*, vol. 123, no. 432, pp. 1–12, 2017, doi: 10.1007/s00339-017-1047-6.

[42] V. V Shvartsman, W. Kleemann, J. Dec, Z. K. Xu, and S. G. Lu, "Diffuse phase transition in $BaTi_{1-x}Sn_xO_3$ ceramics : An intermediate state between ferroelectric and relaxor behavior," *J. Appl. Phys.*, vol. 99, no. 124111, pp. 1–8, 2006, doi: 10.1063/1.2207828.

[43] T. F. Zhang, X. G. Tang, Q. X. Liu, Y. P. Jiang, X. X. Huang, and Q. F. Zhou, "Energy-storage properties and high-temperature dielectric relaxation behaviors of relaxor ferroelectric $Pb(Mg_{1/3}Nb_{2/3})O_3$-$PbTiO_3$ ceramics," *J. Phys. D. Appl. Phys.*, vol. 49, no. 095302, pp. 1–7, 2016, doi: 10.1088/0022-3727/49/9/095302.

[44] Z. Hanani *et al.*, "Phase transitions, energy storage performances and electrocaloric effect of the lead-free $Ba_{0.85}Ca_{0.15}Zr_{0.10}Ti_{0.90}O_3$ ceramic relaxor," *J. Mater. Sci. Mater. Electron.*, vol. 30, no. 7, pp. 6430–6438, 2019, doi: 10.1007/s10854-019-00946-5.

[45] J. Gao *et al.*, "Enhancing dielectric permittivity for energy-storage devices through tricritical phenomenon," *Sci. Rep.*, vol. 7, no. 40916, pp. 1–10, 2017, doi: 10.1038/srep40916.

[46] A. S. Mischenko, Q. Zhang, R. W. Whatmore, J. F. Scott, and N. D. Mathur, "Giant electrocaloric effect in the thin film relaxor ferroelectric 0.9 Pb $Mg_{1/3}Nb_{2/3}O_3$-0.1 $PbTiO_3$ near room temperature," *Appl. Phys. Lett.*, vol. 89, no. 242912, pp. 1–3, 2006, doi: 10.1063/1.2405889.

[47] K. Morimoto, S. Sawai, K. Hisano, and T. Yamamoto, "Simultaneous measurement of specific heat, thermal conductivity, and thermal diffusivity of modified barium titanate ceramics," *Thermochim. Acta*, vol. 442, pp. 14–17, 2006, doi: 10.1016/j.tca.2005.11.020.

[48] X. Wang *et al.*, “Giant electrocaloric effect in lead-free $Ba_{0.94}Ca_{0.06}Ti_{1-x}Sn_xO_3$ ceramics with tunable Curie temperature,” *Appl. Phys. Lett.*, vol. 107, no. 252905, pp. 1–5, 2015, doi: 10.1063/1.4938134.

[49] L. Padurariu, L. Curecheriu, V. Buscaglia, and L. Mitoseriu, “Field-dependent permittivity in nanostructured $BaTiO_3$ ceramics: Modeling and experimental verification,” *Phys. Rev. B - Condens. Matter Mater. Phys.*, vol. 85, no. 224111, pp. 1–9, 2012, doi: 10.1103/PhysRevB.85.224111.

[50] Y. Liu, J. F. Scott, and B. Dkhil, “Direct and indirect measurements on electrocaloric effect: Recent developments and perspectives,” *Appl. Phys. Rev.*, vol. 3, no. 031102, pp. 1–17, 2016, doi: 10.1063/1.4958327.

[51] Z. Kutnjak, B. Rožič, and R. Pirc, “Electrocaloric Effect: Theory, Measurements, and Applications,” *Wiley Encycl. Electr. Electron. Eng.*, pp. 1–19, 2015, doi: 10.1002/047134608X.W8244.

[52] Y. Bai, X. Han, K. Ding, and L. Qiao, “Combined effects of diffuse phase transition and microstructure on the electrocaloric effect in $Ba_{1-x}Sr_xTiO_3$ ceramics,” *Appl. Phys. Lett.*, vol. 103, no. 162902, pp. 1–4, 2013, doi: 10.1063/1.4825266.

[53] E. Defay, S. Crossley, S. Kar-narayan, X. Moya, and N. D. Mathur, “The Electrocaloric Efficiency of Ceramic and Polymer Films,” vol. 25, pp. 3337–3342, 2013, doi: 10.1002/adma.201300606.

[54] T. F. Zhang *et al.*, “Enhanced electrocaloric analysis and energy-storage performance of lanthanum modified lead titanate ceramics for potential solid-state refrigeration applications,” *Sci. Rep.*, vol. 8, no. 1, pp. 1–12, 2018, doi: 10.1038/s41598-017-18810-z.

[55] Y. Zhao, X. Hao, and Q. Zhang, “A giant electrocaloric effect of a $Pb_{0.97}La_{0.02}(Zr_{0.75}Sn_{0.18}Ti_{0.07})O_3$ antiferroelectric thick film at room temperature,” *J. Mater. Chem. C*, vol. 3, no. 8, pp. 1694–1699, 2015, doi: 10.1039/c4tc02381a.

[56] X. Zhang *et al.*, “Large electrocaloric effect in $Ba(Ti_{1-x}Sn_x)O_3$ ceramics over a broad temperature region,” *AIP Adv.*, vol. 5, no. 047134, pp. 1–7, 2015, doi: 10.1063/1.4919096.

[57] C. Molin, T. Richter, and S. E. Gebhardt, “Tailoring electrocaloric properties of $Ba_{1-}$

$_x$Sr$_x$Sn$_y$Ti$_{1-y}$O$_3$ ceramics by compositional modification," *J. Eur. Ceram. Soc.*, vol. 42, no. 1, pp. 140–146, 2022, doi: 10.1016/j.jeurceramsoc.2021.10.003.

[58] M. D. Li *et al.*, "Large Electrocaloric Effect in Lead-free Ba(Hf$_x$Ti$_{1-x}$)O$_3$ Ferroelectric Ceramics for Clean Energy Applications," *ACS Sustain. Chem. Eng.*, vol. 6, no. 7, pp. 8920–8925, 2018, doi: 10.1021/acssuschemeng.8b01277.

[59] Y. Zhao, X. Q. Liu, S. Y. Wu, and X. M. Chen, "Effect of phase transition on electrocaloric effect in Indium substituted $BaTiO_3$ ceramics," *J. Alloys Compd.*, vol. 822, no. 153632, pp. 1–8, 2020, doi: 10.1016/j.jallcom.2019.153632.

**Tables and figures captions**

Table 1: Dielectric parameters obtained from the Santos–Eiras equation of BSTS-x samples at 1 kHz.

Table 2: Comparison of the energy storage performances for well-known lead-free ferroelectric ceramics and the investigated compositions.

Table 3: Comparison of EC characteristics of BSTS-x ceramics developed in this work with typical ferroelectric materials in the literature.

Figure 1: Theoretical and experimental plots of $\varepsilon'$ vs. T for BSTS-x ceramics (x = 0.00, 0.05 and 0.10) at 1 kHz of frequency. The inset depicts the phase diagram of BSTS-x ceramics with Sn doping levels ranging from 0% to 15% [26].

Figure 2: Schematic illustration of the calculation of the energy storage parameters, e.g., x = 0.10, the green area and the red area depict $W_{rec}$ and $W_{loss}$, respectively.

Figure 3: (a-c) Temperature dependence of P – E hysteresis loops of BSTS-x ceramics and (d-f) evolution of polarization vs. temperature of BSTS-x ceramics under various electric fields.

Figure 4: Thermal evolution of (a) recovered energy density, (b) total energy density and (c) energy efficacity of BSTS-x ceramics. (d) Recoverable energy density and energy storage efficiency as a function of Sn-content.

Figure 5: (a-c) Electrocaloric temperature change ($\Delta T_{EC}$) as a function of the temperature at different applied electric fields of BSTS-x ceramics and (d) Electrocaloric temperature change $\Delta T_{EC}$ as a function of the temperature at an applied electric field of 4.4 kV/cm for all samples.

Figure 6: Temperature dependence of the entropy change ($\Delta S$) at various applied electric fields of BSTS-x ceramics (a-c) and an applied electric field of 4.4 kV/cm for all samples (d).

Figure 7: Thermal evolution of direct electrocaloric temperature change ($\Delta T_{EC}$) under different applied electric fields of BSTS-x ceramics. The insets compare direct and indirect EC measurements at the same applied electric field.

Figure 8: Variation of the coefficient of performance COP with composition for BSTS-x ceramics.

Table 1

| Composition x | $\varepsilon'_m$ | $T_m$ $(K)$ | $\Delta$ $(K)$ | $\gamma$ |
|---|---|---|---|---|
| **0.00** | 14670.59 | 368 | 12.41 | 1.01 |
| **0.05** | 18998.53 | 337 | 14.20 | 1.36 |
| **0.10** | 33657.09 | 299 | 15.70 | 1.61 |

Table 2

| Material | $W_{rec}$ (mJ/cm$^3$) | $\eta$ (%) | E(kV/cm) | T(K) | Reference |
|---|---|---|---|---|---|
| **$Ba_{0.90}Sr_{0.10}TiO_3$** | **24** | **85** | **7.4** | **388** | **This work** |
| **$Ba_{0.90}Sr_{0.10}Ti_{0.95}Sn_{0.05}O_3$** | **18** | **92** | **6.8** | **368** | **This work** |
| **$Ba_{0.90}Sr_{0.10}Ti_{0.90}Sn_{0.10}O_3$** | **10** | **89** | **4.4** | **328** | **This work** |
| $BaTi_{0.895}Sn_{0.105}O_3$ | 31 | - | 10 | - | [45] |
| $Ba_{0.85}Ca_{0.15}Zr_{0.10}Ti_{0.90}O_3$ | 14 | 80 | 6.5 | 393 | [44] |
| $Ba_{0.90}Sr_{0.10}TiO_3$ | 144 | - | 33 | 303 | [13] |
| $Ba_{0.70}Sr_{0.30}TiO_3$ | 90 | - | 33 | 303 | [13] |
| $BaTiO_3$ | 71 | 20 | 24 | - | [16] |
| $BaCe_{0.10}Ti_{0.90}O_3$ | 77 | 39 | 24 | - | [16] |
| $BaCe_{0.15}Ti_{0.85}O_3$ | 115 | 65 | 24 | - | [16] |

Table 3

| Material | $T_C$ (K) | $\Delta T_{EC}$ (K) | E (kV/cm) | ξ (K mm/kV) | ΔS (J/Kg.K) | Method | Ref. |
|---|---|---|---|---|---|---|---|
| $Ba_{0.90}Sr_{0.10}TiO_3$ | 360 | 0.61 | 33 | 0.18 | 0.70 | Indirect | [13] |
| **$Ba_{0.90}Sr_{0.10}TiO_3$** | **368** | **0.545** | **7.4** | **0.73** | **0.57** | **Indirect** | **This work** |
| **$Ba_{0.90}Sr_{0.10}TiO_3$** | **366** | **0.522** | **7.4** | **0.70** | **-** | **Direct** | **This work** |
| **$Ba_{0.90}Sr_{0.10}Ti_{0.95}Sn_{0.05}O_3$** | **341** | **0.207** | **6.8** | **0.30** | **0.26** | **Indirect** | **This work** |
| **$Ba_{0.90}Sr_{0.10}Ti_{0.95}Sn_{0.05}O_3$** | **338** | **0.463** | **16.0** | **0.29** | **-** | **Direct** | **This work** |
| **$Ba_{0.90}Sr_{0.10}Ti_{0.90}Sn_{0.10}O_3$** | **300** | **0.101** | **4.4** | **0.23** | **0.14** | **Indirect** | **This work** |
| **$Ba_{0.90}Sr_{0.10}Ti_{0.90}Sn_{0.10}O_3$** | **302** | **0.190** | **8.5** | **0.22** | **-** | **Direct** | **This work** |
| $Ba_{0.80}Sr_{0.20}TiO_3$ | 338 | 0.83 | 33 | 0.25 | 1.0 | Indirect | [13] |
| $Ba_{0.65}Sr_{0.35}TiO_3$ | 296 | 0.42 | 20 | 0.21 | - | Indirect | [52] |
| $BaTi_{0.90}Sn_{0.10}O_3$ | 335 | 0.207 | 10 | 0.20 | 0.27 | Indirect | [56] |
| $BaTi_{0.89}Sn_{0.11}O_3$ | 317 | 0.63 | 20 | 0.31 | 0.87 | Direct | [24] |
| $Ba_{0.85}Sr_{0.15}Sn_{0.05}Ti_{0.95}O_3$ | 320 | 1.44 | 30 | 0.48 | - | Direct | [25] |
| $Ba_{0.94}Ca_{0.06}Ti_{0.95}Sn_{0.05}O_3$ | 358 | 0.59 | 20 | 0.30 | - | Indirect | [48] |
| $Ba_{0.94}Ca_{0.06}Ti_{0.90}Sn_{0.10}O_3$ | 320 | 0.55 | 20 | 0.28 | - | Indirect | [48] |
| $Ba_{0.96}La_{0.04}TiO_3$ | 303 | 1.81 | 60 | 0.30 | - | Indirect | [12] |
| $Ba_{0.82}Sr_{0.18}Sn_{0.065}Ti_{0.935}O_3$ | 304 | 0.49 | 20 | 0.25 | - | Direct | [57] |
| $Ba(Hf_{0.05}Ti_{0.95})O_3$ | 349 | 1.21 | 50 | 0.24 | - | Indirect | [58] |
| $Ba(Ti_{0.99}In_{0.01})O_{2.995}$ | 290 | 0.22 | 10 | 0.22 | 0.39 | Indirect | [59] |
| $BaTiO_3$ (single-crystal) | 402 | 0.9 | 12 | 0.75 | - | Direct | [4] |
| $PbMg_{1/3}Nb_{2/3}O_3$ | 340 | 2.5 | 90 | 0.27 | 2.5 | Direct | [15] |

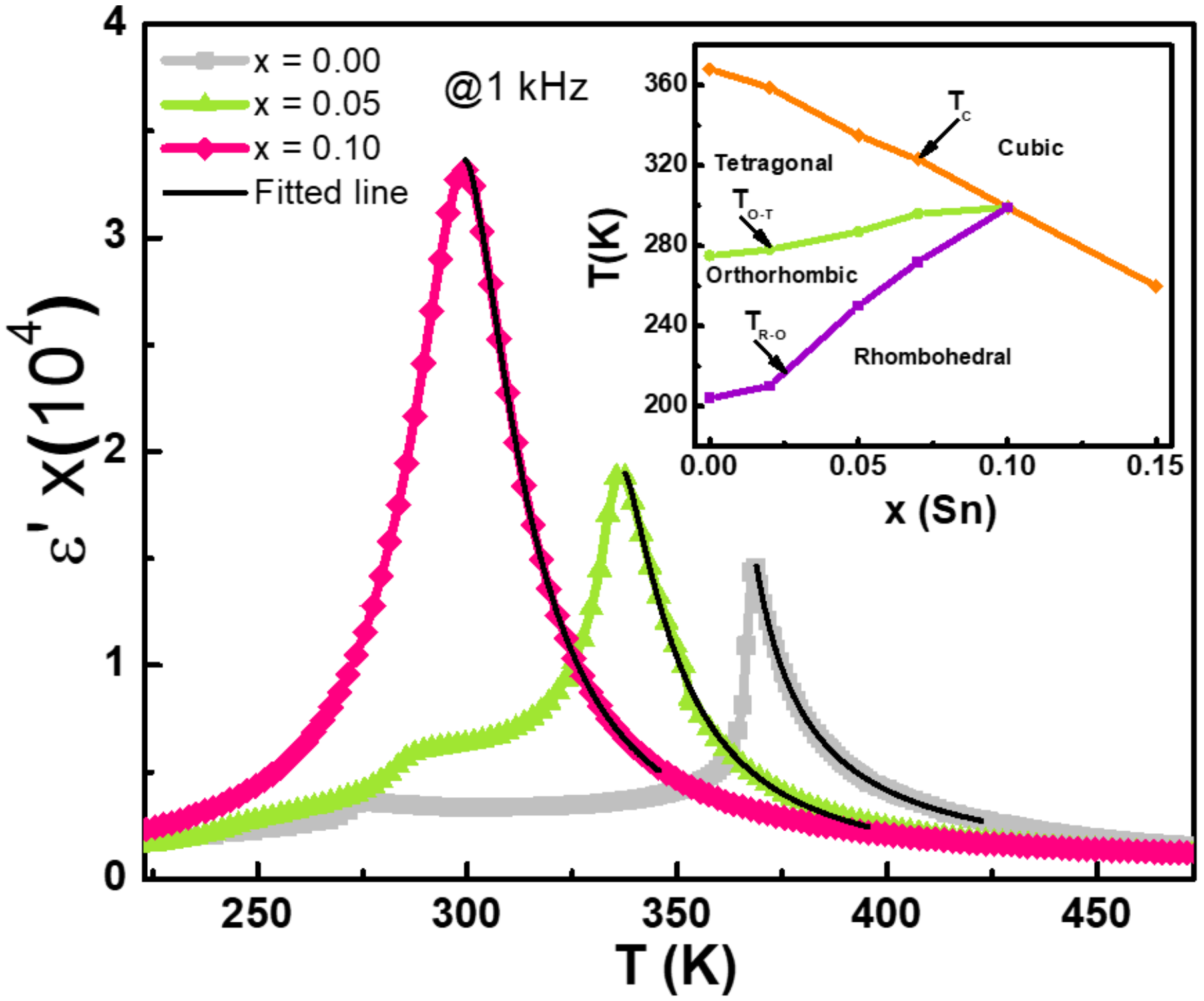

Figure 1

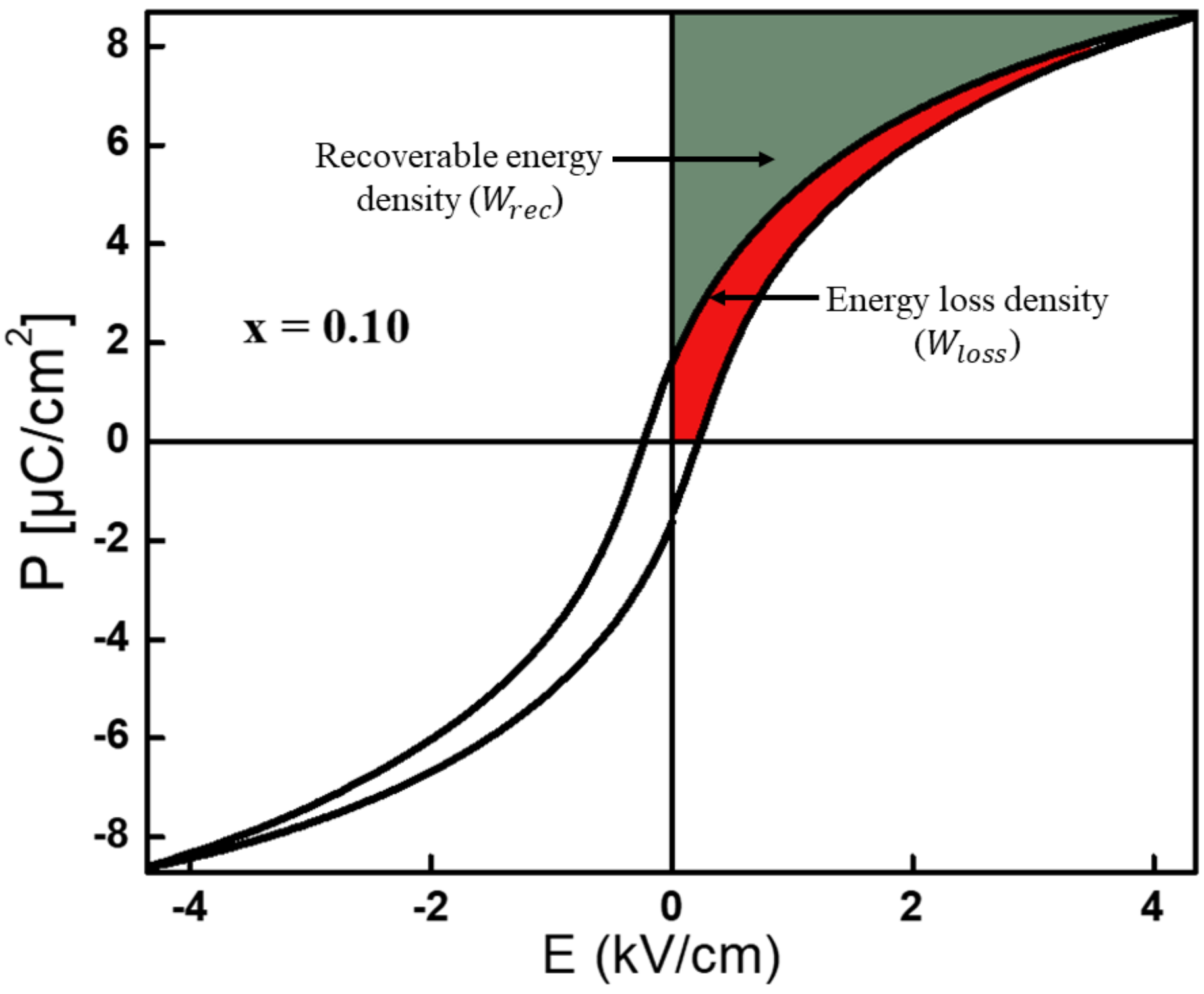

Figure 2

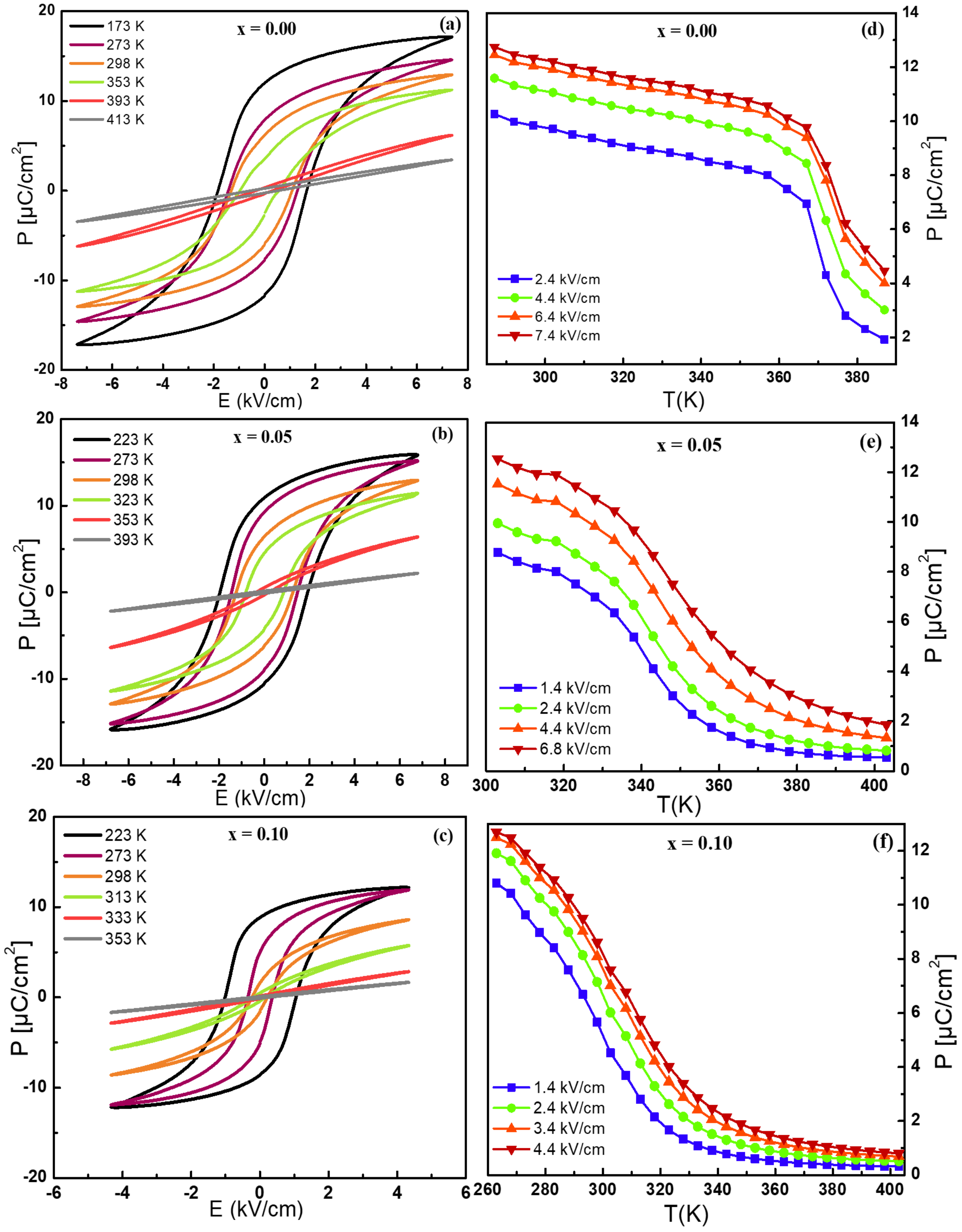

Figure 3

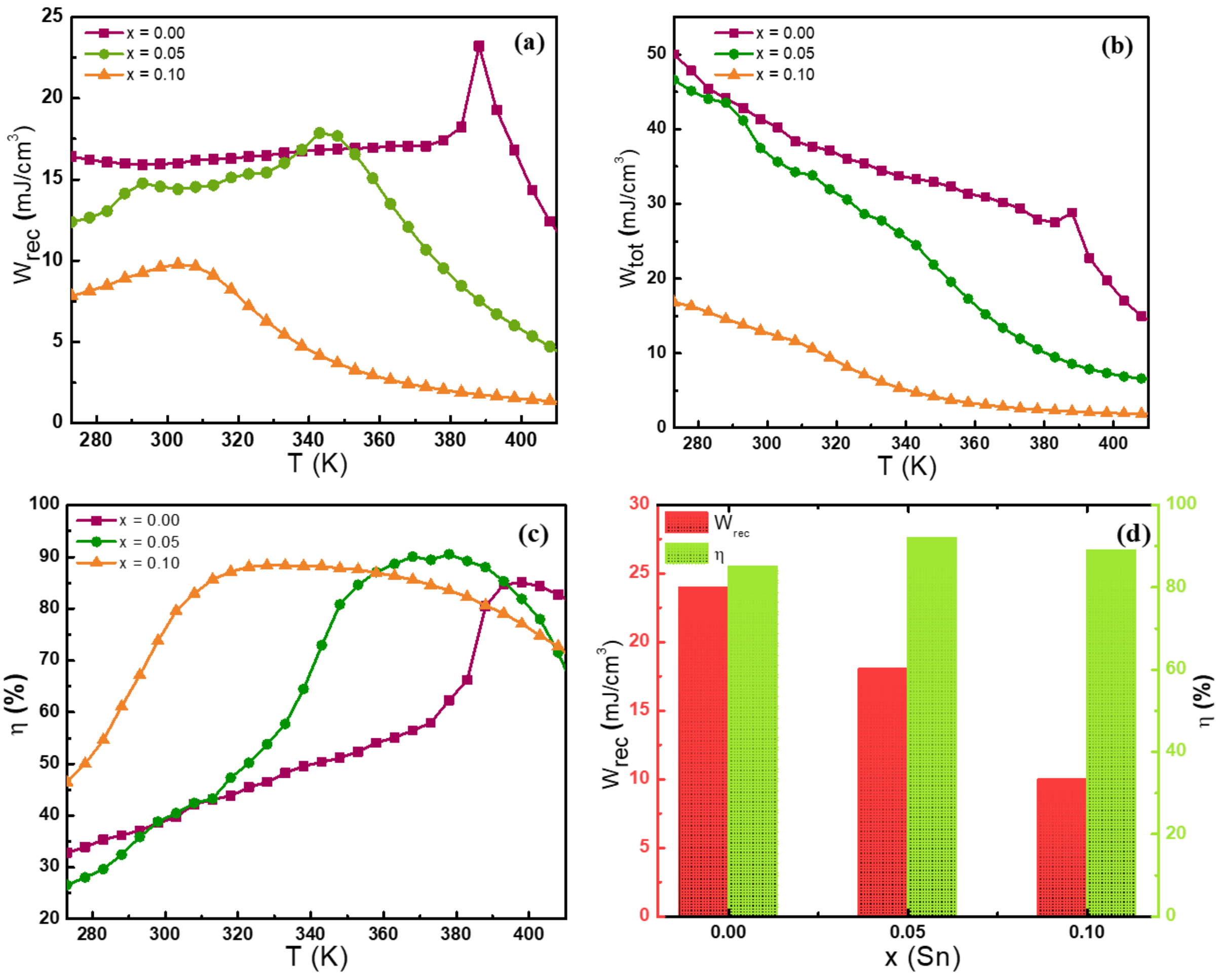

Figure 4

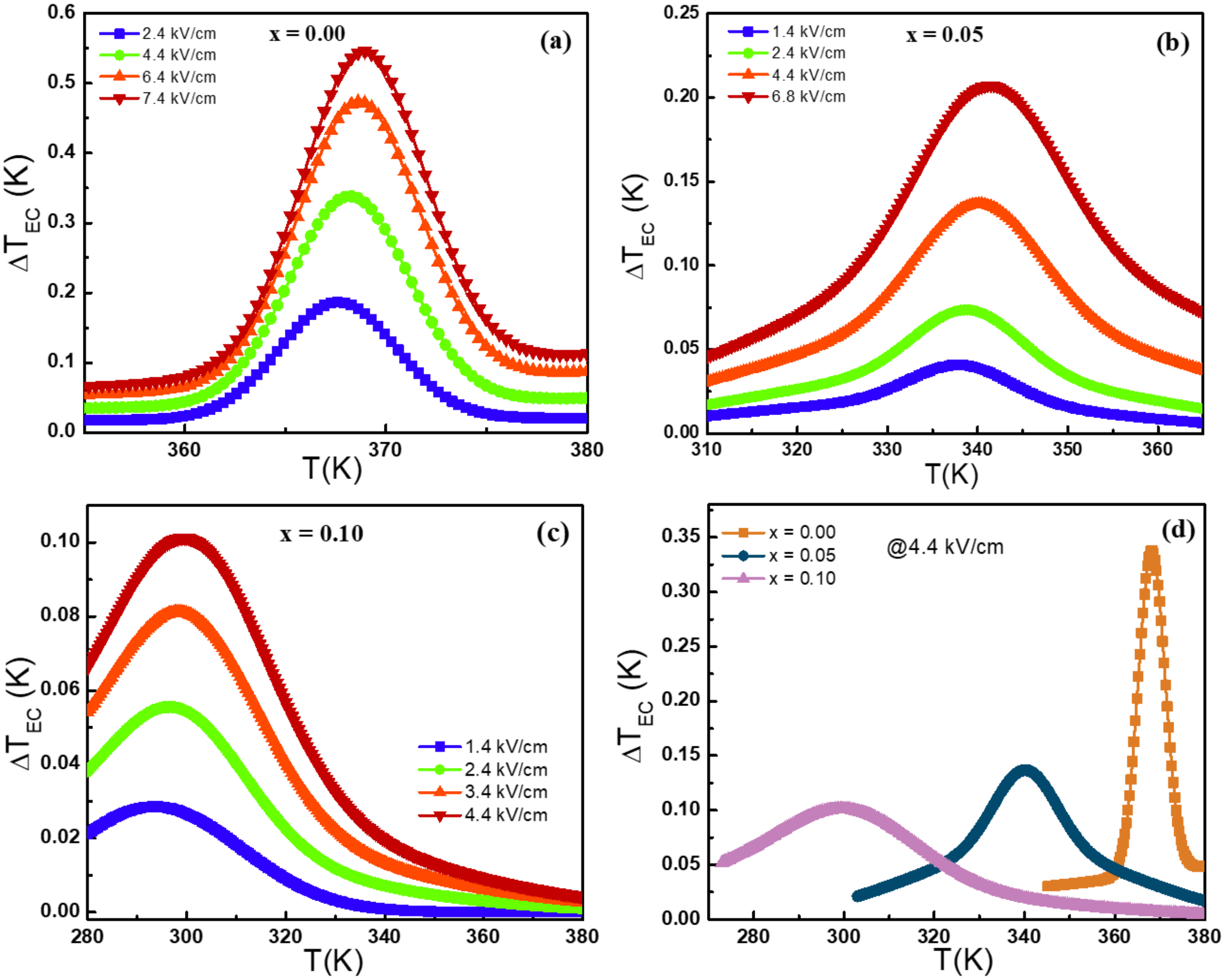

Figure 5

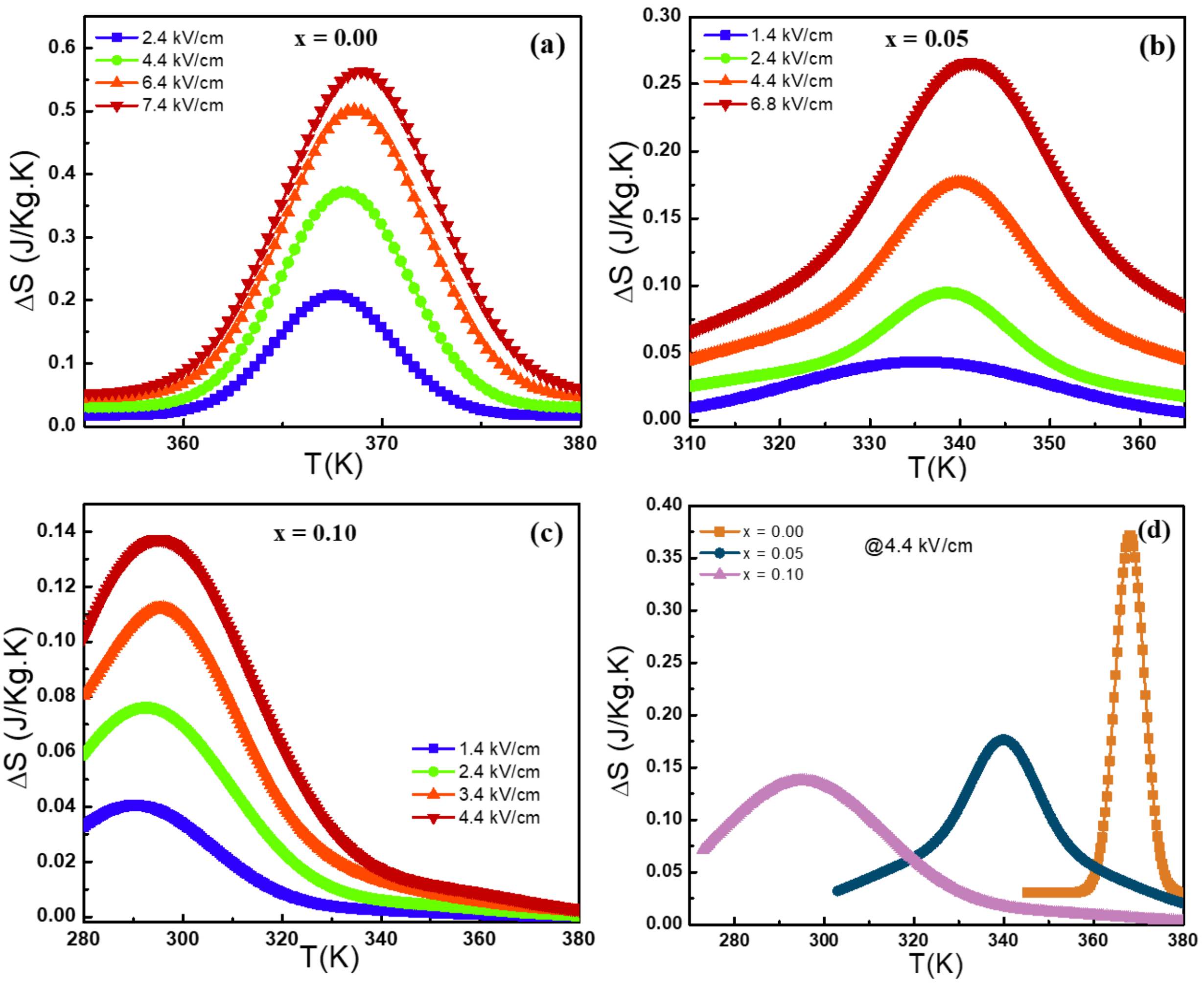

Figure 6

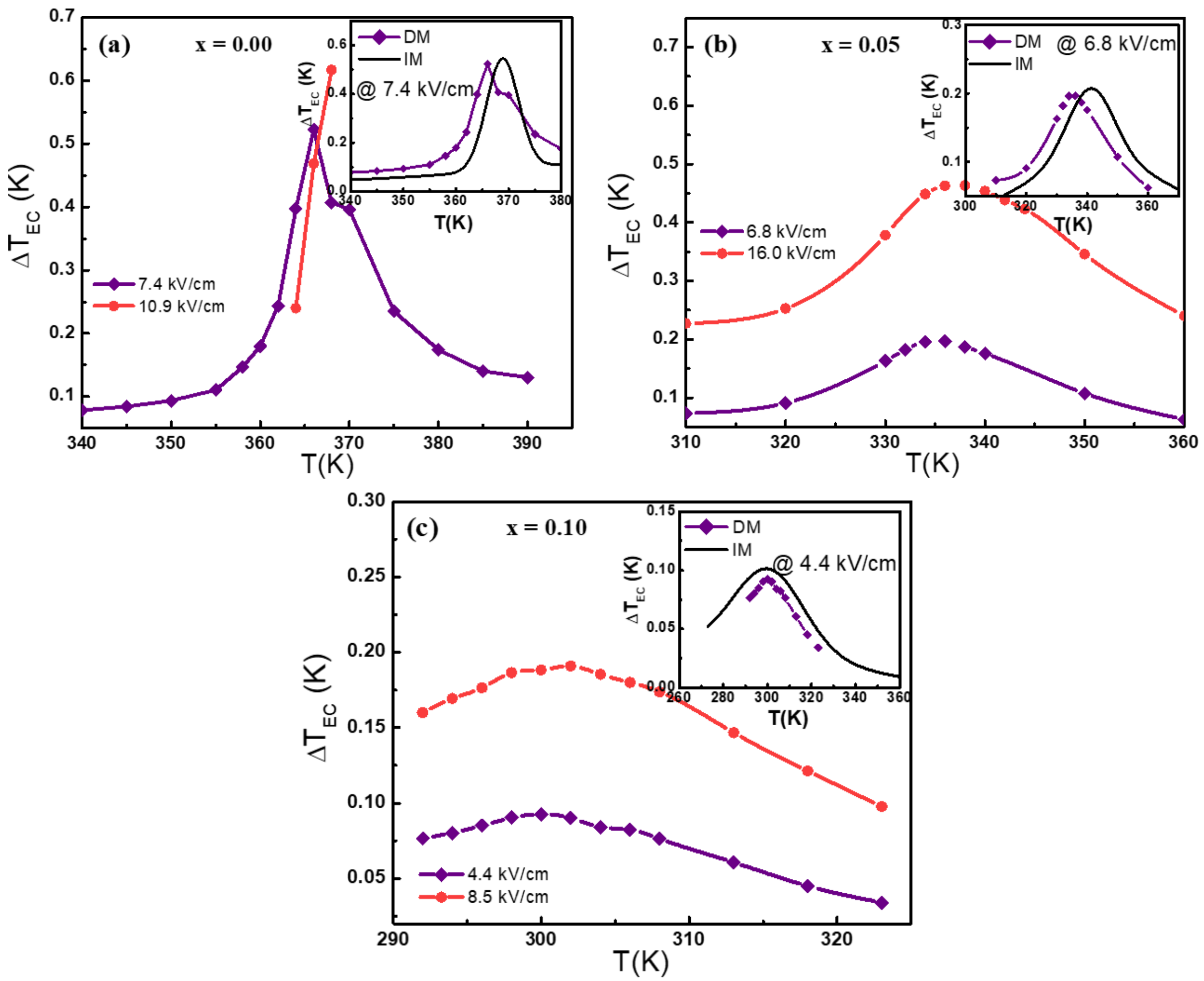

Figure 7

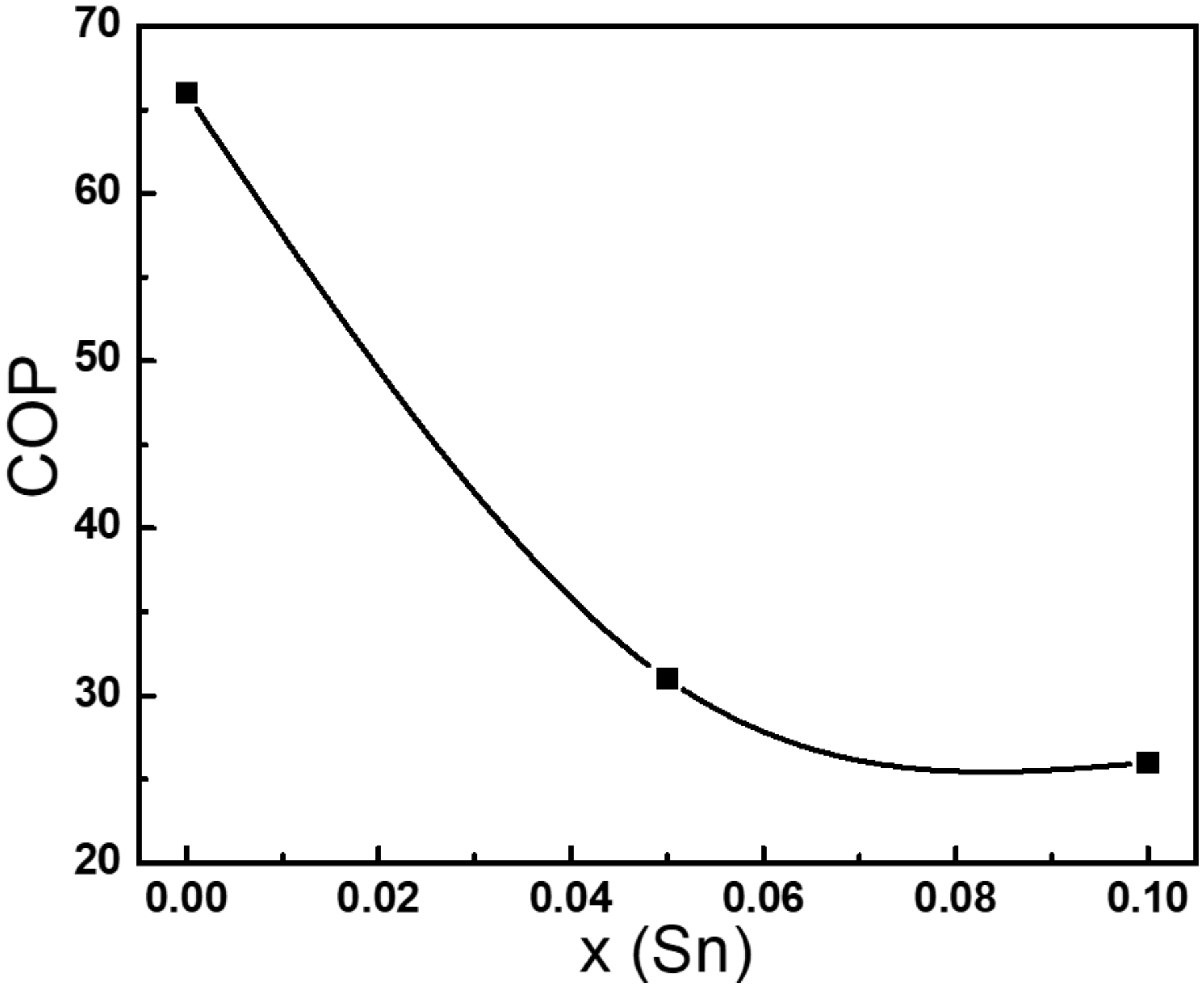

Figure 8